\documentclass[letterpaper, 10pt]{article}
\usepackage{graphicx}
\usepackage{float}
\usepackage[inner=0.75in, outer=0.75in, top=0.75in, bottom=1in, paperheight=11in, paperwidth=8.5in]{geometry}

\usepackage{array}
\usepackage{rotating}
\usepackage{multirow}

\usepackage{mathtools}
\usepackage{shuffle}
\usepackage[table, svgnames, dvipsnames,xcdraw]{xcolor}
\usepackage[square, numbers, comma, sort&compress]{natbib}
\usepackage{verbatim}
\usepackage{bm}
\usepackage{amsmath}
\usepackage{amsfonts}
\usepackage{amssymb}
\usepackage{amsthm}

\usepackage{epstopdf}
\usepackage[utf8]{inputenc}
\usepackage[english]{babel}
\usepackage{color}
\usepackage{forest}

\usepackage{pdfpages}
\usepackage{pgfplots}
\usepackage{pgfplotstable}
\usepgfplotslibrary{patchplots}
\pgfplotsset{compat=1.15}

\usepackage{microtype}
\usepackage{url}
\usepackage{subfiles}
\usepackage{caption}
\usepackage{subcaption}
\usepackage{parskip}
\usepackage[section]{placeins}

\usepackage{tikz}
\usepackage{tikz-cd}
\usetikzlibrary{arrows,calc,shapes,decorations.pathreplacing}
\usetikzlibrary{decorations.markings}
\usetikzlibrary{trees,positioning,fit,shapes}
\usetikzlibrary{backgrounds}
\usetikzlibrary{patterns, positioning, arrows}

\usepackage{framed}

\colorlet{shadecolor}{blue!20}

\usepackage[obeyFinal,textwidth=0.55in]{todonotes}
\usepackage[hidelinks]{hyperref}
\usepackage[capitalize]{cleveref}

\usepackage{enumitem}

\theoremstyle{plain}

\theoremstyle{definition}

\begin{document}	

  \title{ \bf Bayesian graph neural networks for strain--based crack localization }
  \author{C.\ Mylonas$^1$ ~ G.\ Tsialiamanis$^2$ ~ K.\ Worden$^2$ ~ E.\ N.\ Chatzi$^1$ \\ 
     $^1$Institute of Structural Engineering, ETH Z\"{u}rich, \\ 
     Stefano-Franscini-Platz 5, CH-8093 Z\"{u}rich, Switzerland\\
     $^2$ Dynamics Research Group, Department of Mechanical Engineering, \\University of Sheffield Mappin Street, Sheffield S1 3JD, UK }
\date{}
\maketitle

\newcommand\copyrighttext{%
  \footnotesize This is a preprint of a work submitted for possible publication on conference proceedings. Copyright may be transferred without notice, after which this version may no longer be accessible. }
\newcommand\copyrightnotice{%
\begin{tikzpicture}[remember picture,overlay]
\node[anchor=south,yshift=10pt] at (current page.south) {\fbox{\parbox{\dimexpr\textwidth-\fboxsep-\fboxrule\relax}{\copyrighttext}}};
\end{tikzpicture}%
}

\section*{Abstract}

A common shortcoming of vibration-based damage localization techniques is that localized damages, i.e. small
cracks, have a limited influence on the spectral characteristics of a structure. In contrast, even the smallest of defects, under particular loading conditions, cause localized
strain concentrations with predictable spatial configuration. However, the effect of a small defect on strain decays quickly with distance from the defect, making strain-based localization rather challenging. In this work, an attempt is made to approximate, in a fully data-driven manner, the posterior distribution of a crack location, given arbitrary dynamic strain measurements at arbitrary discrete locations on a structure. The proposed technique leverages Graph Neural Networks (GNNs) and recent developments in scalable learning for Bayesian neural networks. The technique is demonstrated on the problem of inferring the position of an unknown crack via patterns of dynamic strain field measurements at discrete locations. The dataset consists of simulations of a hollow tube under random time-dependent excitations with randomly sampled crack geometry and orientation. 

\textbf{Keywords: GraphNets, Strain-based crack localization, Bayesian Graph Neural Network}

\copyrightnotice
\section{Introduction}

When the region around a defect is loaded, strain concentration occurs. For different local loading conditions, depending on the geometric configuration 
of the defect, different strain concentration patterns arise.
The technique proposed in this work, allows for strain-based localization of cracks with learned, rather than hand-engineered, features. 
In contrast to the approaches based on 2D Convolutional Neural Networks (CNNs) already in the literature \cite{gulgec2019convolutional}, 
the approach presented herein can effectively deal with arbitrary (potentially sparse) strain sensor positions, and not only sensing 
positions arranged in a grid as CNN-based computation dictates. This feature is achieved by employing a message-passing 
Graph Neural Network (GNN) \cite{mpnn, battaglia2018relational}. The GNN is used for recognising spatial strain patterns,
and learns these patterns directly from data. These patterns depend non-linearly on the sensor position 
relative to the crack and the sensor position in the structure.

In essence, as CNNs perform local aggregation defined by a spatial 2D grid, message-passing graph neural networks allow for 
local aggregation on the neighborhood of the nodes of an arbitrary graph.
\begin{figure}[ht!]
	\center
	\includegraphics[width=\textwidth]{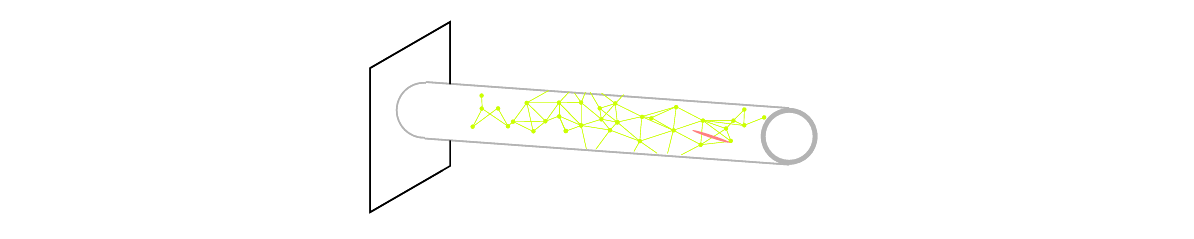}
  \caption{Schematic of the construction of a strain sensing graph. Each node corresponds to a full strain-sensor. 
  The neighborhood of each sensor is defined by the sensors closer to it. An ellipsoidal crack is depicted in red.}
	\label{fig:strainsens}
\end{figure}
In contrast to other works in the literature \cite{agathos2018multiple, agathos2020parametrized}, a model that predicts directly the 
crack position from data is learned and no optimization is performed during prediction time\footnote{This work formed part of the PhD thesis of the first author \cite{mylonas2021machine}. }.

\paragraph{Training Data And Strain-PCA contrasting:}
In this study, the dynamics of 450 hollow tubes with randomly-positioned and rotated ellipsoidal defects were 
simulated with the Finite Element Method under random white-noise excitations in three dimensions, different for every run. 
The full strain tensors on a $150 \times 150$ grid of values on the surface of the tube were extracted. All 
simulated tubes are 10m in length and 1m in diameter.

\begin{figure}[ht!]
	\center
  \includegraphics[width=\textwidth]{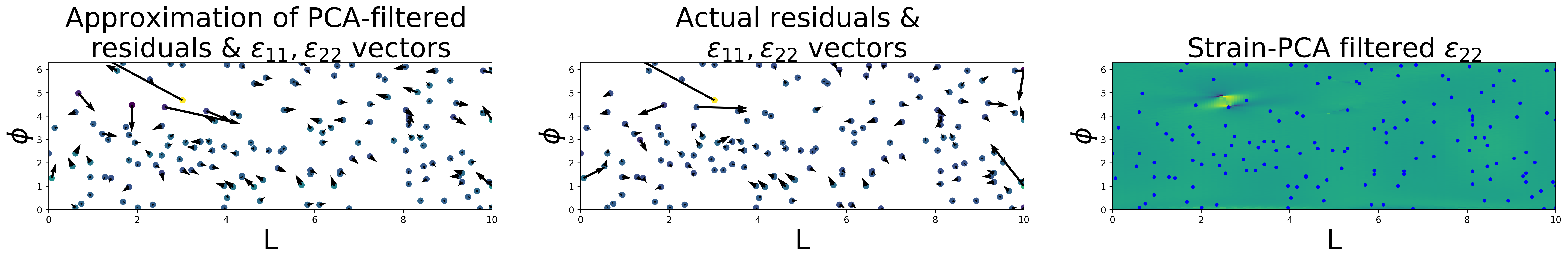}
  \caption{Contrasting sparse measurements using strain field PCA. The positions on the plots correspond to length-wise and angular cylinder coordinates. The left-most plot is an approximation of the defect-dependent residual based on the projection on a limited number of sensors with $\bar \Psi_{\varepsilon}$. The middle plot is the actual residual based on a projection using the dense measurements and full PCA components $\Psi_\varepsilon$. The rightmost plot is the full residual strain field for $\varepsilon_{22}$ for a time-instant where $\varepsilon_{22}$ reveals the position of the strain.}
	\label{fig:svdfilt}
\end{figure}

A basis is first obtained from a Principal Components Analysis (PCA) of strain snapshots of a crack-free structure.
In the presented results, 50 components of the PCA basis are retained (variance explained $88.3\%$).
A separate PCA for each component of the strain tensor and the strain-invariants is learned.
The strain fields corresponding to structures that contain a defect are pre-processed by projecting their strain fields to the corresponding
PCA-derived bases. The projected strains are removed from the observed strains. Assuming a linear superposition of the strain related to the defect
and the strain of the defect-free structure, the strain associated with the defect should appear in the residuals of the projection. Indeed, this effect is demonstrated in the right-most plot of \autoref{fig:svdfilt}.
During prediction, only a sparse set of sensors is expected to be available. Therefore, the estimates of the strain 
field are going to be less accurate compared to the case where the full strain field is available. The projection using a sparse mask of strain sensors,
is equivalent to performing a simple linear regression with the PCA bases. 
The PCA approximation of $\hat \varepsilon$, for a strain component $\varepsilon$ based on a sparse set of sensors at $i$ positions $S : \{s_1, \dots s_i\}$, is computed from
\newcommand\sparseassign{\stackrel{\mathclap{\normalfont\mbox{S}}}{\leftarrow}}
\begin{align*}
  \bar \Psi_{\varepsilon} & \sparseassign \Psi_{\varepsilon} \\ 
  \hat c &= (\bar \Psi_\varepsilon^T\bar\Psi_{\varepsilon})^{-1} \bar \Psi_{\varepsilon}^T \bar \varepsilon \\
  \hat \varepsilon &= \bar \Psi_\varepsilon \hat c \\ 
  r^{PCA}_\varepsilon &= \hat \varepsilon - \varepsilon \\
\end{align*}
where $\Psi_{\varepsilon}$ is the matrix of PCA components, $\bar \Psi_{\varepsilon}$ is the matrix of PCA components, sparsified at the sensor positions, $\hat c$ is a vector of 
coefficients and $r^{PCA}_\varepsilon$ is the PCA-computed residual for that strain component. 
In what follows, for notational convenience the PCA-derived strain residuals $r^{PCA}_{\varepsilon_{ij}} $ are referred to simply as strains $\varepsilon_{ij}$.

A representative example of the quality of the approximation for the investigated structure, and for the crack geometries used 
is depicted in \autoref{fig:svdfilt}.

\section{Message-Passing Graph Neural Networks}
A directed attributed graph $G : ( V, E )$, (referred to in the following simply as ``graph'') is defined as a set of vertices (nodes)
$V: \{ \mathbf{v}_1, \dots \mathbf{v}_{N^n}\}$ and a set of edges $E : \{ (\mathbf{e_1}, s_{1}, r_{1}), \cdots (\mathbf{e}_{N^e}, s_{N^e}, r_{N^e} \} $. 
Nodes contain attributes, and edges are described by an attribute $\mathbf{e}_i$, and sender $s_{i}$ and receiver $ r_{i}$ indices.
Indices $s_i, r_i$ are integers representing the sender and receiver nodes. \textit{Attributes} may be discrete or continuous vectors or tensors.
In \cite{battaglia2018relational} the \textit{GraphNet} (GN) computational block is proposed, which is a generalisation of several approaches to computation using attributed graphs, 
with a focus on implementing \textit{inductive biases}. GNs are algorithms that operate on graphs and return graphs. 
The attributes of the vertices and edges, parametrize how the entities in the graph interact. 
In addition, graph-level (global) attributes may exist, which depend on the aggregation of all vertex and node information.
More concretely, for the problem considered, the strain sensors have the role of vertices, their relative positions are 
edge features (which dictate which nodes will interact to yield a prediction), and the position and orientation of the crack is 
the global variable which needs to be inferred from the sensor observations.

A GraphNet computational block consists of
\begin{itemize}
  \item an edge function $f^e : \mathbf{e} \rightarrow \mathbf{e}'$ which parametrizes what information is passed along an edge ; 
  \item an edge message aggregation function $\rho^{e\rightarrow v}$, which is a permutation invariant function that summarizes the messages to be passed according to the graph connectivity;
  \item a vertex (node) function that transforms the vertex attributes $f^v : \mathbf{v} \rightarrow \mathbf{v}'$, optionally using the aggregated incoming messages, the node attributes and the global attributes, and finally;
  \item a global function $f^u : \mathbf{u} \rightarrow \mathbf{u}'$ ;
which uses global attributes. Node and edge aggregated information
which are aggregated by two additional aggregators:
    \item the edge-to-global aggregator $\rho^{e \rightarrow u}$ and 
    \item the node-to-global aggregator $\rho^{v \rightarrow u}$. 
\end{itemize}

\begin{figure}[ht!]
\center
  \includegraphics[width=\textwidth]{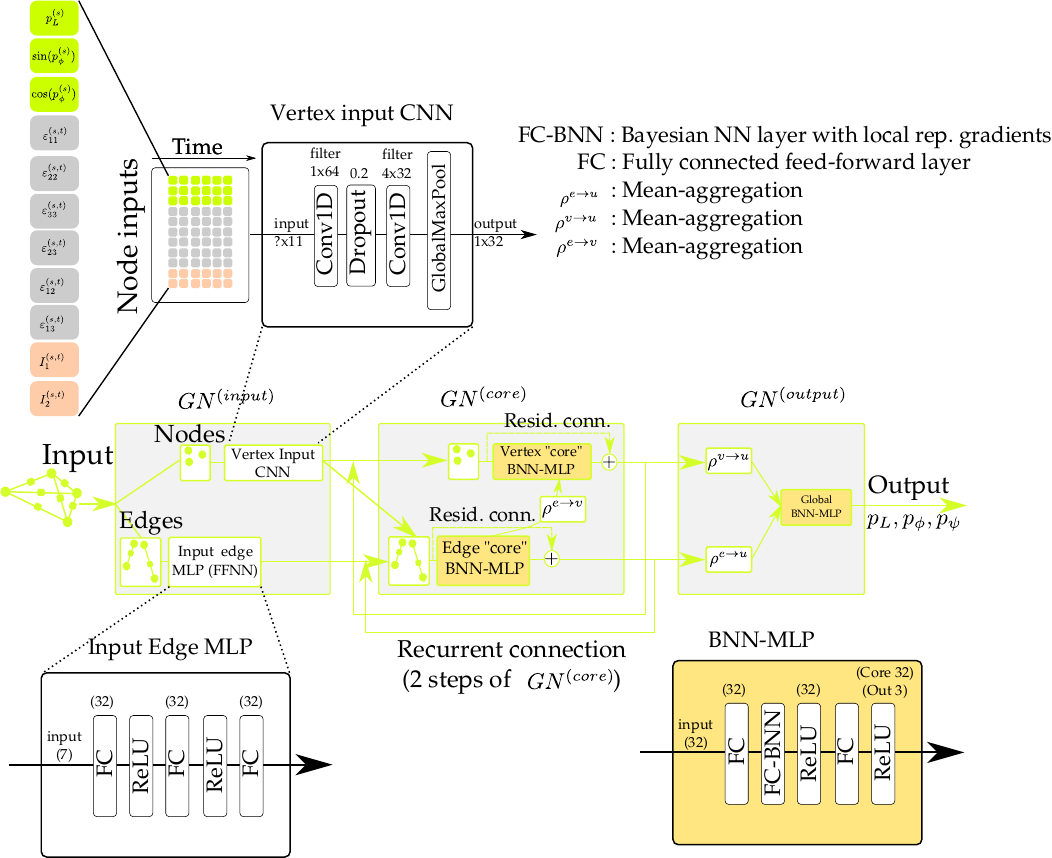}
  \caption{The computational graph used. The $GN^{(core)}$ block is repeatedly applied in order to allow for message passing between a larger neighborhood for each node.}
  \label{fig:compgraph}
\end{figure}

The interaction of vertices in GraphNets occurs through a process that resembles \textit{message passing} \cite{mpnn}, with a 
learned message passing function. In addition to the message-passing function, transformations of the edge and node attributes are performed, 
for instance by a Multi-Layer Perceptron (MLP). 
Graph Neural Networks (GNNs) are GNs that use MLPs for one or several of the aforementioned functions.
The reader is referred to \cite{battaglia2018relational} for a more detailed and complete exposition on GraphNets.
A custom TensorFlow-based GraphNets library was used \footnote{ Code can be accessed at: \href{https://github.com/mylonasc/tf-gnns}{https://github.com/mylonasc/tf-gnns}}. 
The actual computation graph used in this work is given in \autoref{fig:compgraph}. 
The construction of the graphs and the choices for the node, edge, global and aggregation functions is, in general, problem specific.
In the following, the feature construction and some architectural choices are detailed.

\begin{figure}[ht!]
	\center
	\includegraphics[width = \textwidth]{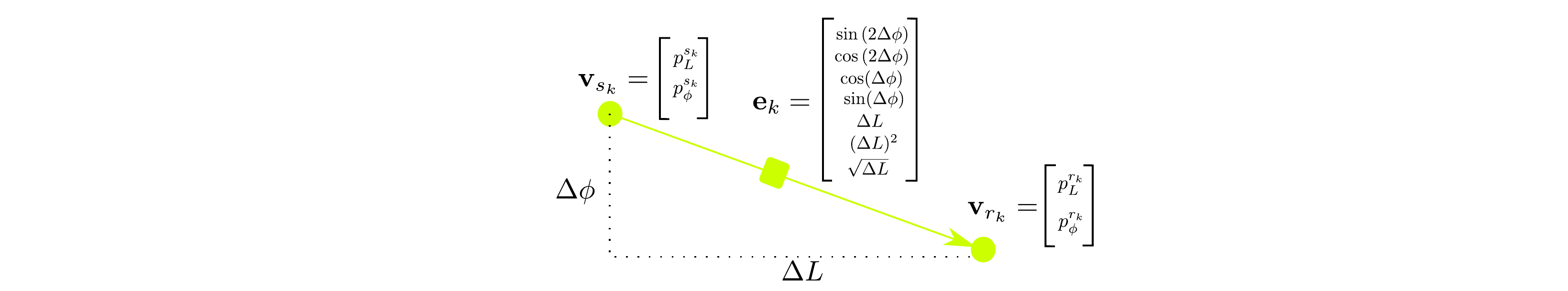}
  \caption{Edge features parametrize the manner in which the different nodes in the GNN exchange information. To that end, non linear relative geometric information is included in the edge function.}
	\label{fig:edgefeatures}
\end{figure}

\paragraph{Node feature construction:}
Since the strain-input of the network is a time-series, it is natural to process it with a 1D convolutional neural network.
The raw strain time-series is augmented by the two first rotational strain invariants, which read,
\begin{align}
  I_1(t, \mathbf{p_i}) &= \varepsilon_{11}(t, \mathbf{p_i}) + \varepsilon_{22}(t, \mathbf{p_i}) + \varepsilon_{33}(t, \mathbf{p_i})\\
  I_2(t,\mathbf{p_i}) &= \varepsilon_{12}(t, \mathbf{p_i})^2 + \varepsilon_{23}(t, \mathbf{p_i})^2 + \varepsilon_{13}(t, \mathbf{p_i})^2 -\\
  & \quad \varepsilon_{11}(t, \mathbf{p_i})\varepsilon_{22}(t, \mathbf{p_i}) -\varepsilon_{22}(t, \mathbf{p_i})\varepsilon_{33}(t, \mathbf{p_i}) - \varepsilon_{33}(t, \mathbf{p_i})\varepsilon_{11}(t, \mathbf{p_i}) 
\end{align}
where $\mathbf{p_i} = [p_L, p_{\phi}]$ is the position of sensor $i$.
The position of the sensors along the length of the tube and the angular dimension, is appended as an additional filter dimension. The angular position $p_\phi$ is encoded as 
$sin(p_\phi),cos(p_\phi)$.
\paragraph{Edge features:}
In order to localize the crack, information between several nearby sensors may need to be taken into account.
Groups of sensors should collectively detect patterns, parametrized by their relative position. This is implemented by endowing the edges of 
the created graph with features that encode this information. Denoting $s_i$ and $r_i$ the sender and receiver node, the features given in \autoref{fig:edgefeatures} 
are used. 
The 1D convolutional network captures temporal variations whereas the GNN should capture parameterized spatial patterns that are associated with the existence of a crack.
\paragraph{Construction of the strain-sensor graph:}
Since local features of the strain field are more informative for the problem, 
the spatial proximities of the randomly-placed strain sensors are used to define the graphs employed in the computations, as depicted in \autoref{fig:strainsens}. 
For each sensor graph, a set of 150 strain sensors are used ($\frac{150}{10\cdot2\cdot\pi\cdot0.5m^2} \approx 4.8[sensors]/[m^2]$). Sensors at a distance less than 10cm from the crack 
are removed to avoid detecting the crack only from the local strain concentration magnitude, which renders the problem much less challenging. Different and unique strain sensor graphs and strain sensor positions were used for each training and testing input. In essense, the crack positions, the loading conditions and the strain sensor configurations encountered at testing, are never encountered during training. The edges in the network are bi-directional, i.e. for each directional edge from node $i$ to node $j$ there is a directional edge from node $j$ to node $i$.

\paragraph{Local reparametrization and uncertainty quantification}
Since, in general, the defect localization problem may be ill-posed\footnote{Consider the cases where the structure is unloaded, or when the structure is loaded in a manner that does not deform the defect. In these two extreme cases, the problem is not solvable. Moreover, there are multiple material configurations that yield the same discrete strain observations for the same loading.}, the uncertainty in the predictions and learned model needs to be represented.
Recent scalable approaches to representing model uncertainty in deep learning include Monte-Carlo Dropout \cite{gal2016dropout} 
and variational inference with stochastic gradient descent \cite{kingma2015variational, bayesbybackprop}. 
Variational inference relies on an approximation of the maximum likelihood objective, 
namely the Evidence Lower Bound (ELBO). In variational techniques, a distribution over the parameters of network layers is inferred from data.
The local reparametrization from \cite{kingma2015variational} is used, since 1D regression experiments showed faster 
and more consistent training. Recent literature for Bayesian GNNs \cite{lamb2020bayesian} supports these observations. 

\subsection{Model architecture}
The node inputs are passed through a 1D-CNN that acts as the node function of the input GNN $GN^{(input)}$.
A simple 2-layer CNN with global max-pooling was chosen, as it is expected that the dynamics play a minor role in identifying the position of the crack from strain. 
Simple two-layer ReLU Multi-Layer Perceptrons (MLPs) were used as an edge function for the input GNN. No message passing is performed in the input GNN (Graph-independent GNN block).
Both $GN^{(input)}$ edge function outputs and $GN^{(input)}$ node-function outputs have a size of 32 units. Networks with fewer units also demonstrated relatively good performance.
Following the input GNN, two full GraphNet (core) layers are applied. Residual connections were used for all MLPs involved in $GN^{(core)}$. 
In the core MLP, instead of the simple two-layer ReLU MLPs, for both the edge and node MLPs an MLP that contains a Bayesian Neural Network layer (BNN layer) is used. 
The \texttt{tensorflow-distributions} implementation \cite{dillon2017tensorflow} of the local-reparametrization layer \cite{kingma2015variational} was used.
The inputs are first passed to a fully-connected layer, a single Bayesian Neural Network (BNN) layer with ReLU activation is then applied, and the MLPs conclude 
with dense layers with no activations. A mean-aggregator was used for the edge-to-node aggregators ($\rho^{e \rightarrow v}$). 

In the final computation step, a graph-to-global layer is employed ($GN^{(output)}$), which directly returns estimates on the position of the crack. 
 The nodes and edges output from the $GN^{(core)}$ are mean-aggregated and concatenated. The same type of BNN-MLP used in the core network is used, with different 
 output of three units, corresponding to $p_\phi, p_L, p_\psi$, which are the crack parameters.

Some preliminary experiments were performed on the performance of the proposed architecture for different choices of hyper-parameters.
Experiments were performed with more $GN^{(core)}$ steps (message-passing through repeated application of the $GN^{(core)}$ network) with and without BNN layers, and it was found 
that two and three layers offered comparable results. More than three $GN^{(core)}$ steps did not offer any improvements. 
Networks with a single message-passing step, regardless of the size of the involved MLPs, performed badly. This observation provides 
some preliminary evidence that aggregating the information from several sensors is important for the crack localization problem.
\paragraph{Training loss and train-test split}
Denoting as $G^{in}$ the input graphs, as $\mathbf{p} : [p_\phi, p_L,p_\psi] $ the corresponding crack positions, $\mathcal{D} : \{ G^{in}, \mathbf{p} \}$ the dataset, as $p(w)$ 
a prior distribution over the parameters of the network, the loss function (ELBO) reads
\begin{align*}
  \mathcal{L(\phi , \mathbf{w}, \mathcal{D} )} = -D_{KL}(q_\phi(\mathbf{w}) || p(\mathbf{w})) + L_{\mathcal{D}}(\phi)
\end{align*}
where $\mathbf{w}$ are all the parameters of the associated MLPs, $\phi$ are the parameters of the variational 
approximation to the posterior $q_{\phi}(\mathbf{w})$, $p(\mathbf{w})$ is a prior distribution (a factored Gaussian).
Moreover, $D_{KL}(q_\phi(\mathbf{w}) || p(\mathbf{w}))$ is the \textit{Kullback-Leibler} divergence, and $L_{\mathcal{D}}$ is 
\begin{equation*}
  L_{\mathcal D}(\phi) = \sum_{(G^{in},\mathbf{p}) \in \mathcal D} E_{q_{\phi}(\mathbf{w})}[\log p(G^{in} | \mathbf{p},\mathbf{w})]
\end{equation*}
which is the \textit{expected negative log-likelihood}. In addition, we are assuming $p(\mathbf{p}) \sim \mathcal{N}(\mu_\mathbf{p}, \sigma^2_\mathbf{p})$, and $q_{\phi}(\mathbf{w}) \sim \mathcal{N}(\mu_\mathbf{w}, \sigma^2_\mathbf{w})$. Low variance stochastic gradient estimators exist for the loss function \cite{kingma2015variational, vaepaper}.


350 experiments were used as training set and the remaining 100 as testing set. The loss on the testing set was monitored and 
early stopping was applied after 50 epochs. Random segments of 150 timesteps from the 401 that are available (0.5 seconds) were used from each experiment of the training set, 
different at each epoch. Interestingly, the network did not suffer from over-fitting, as evaluated in the test-set of unseen sensor configurations and crack positions,
even though the number of experiments in the raw dataset number is relatively small (350).
This effect is probably a result of the combination of the small networks used and the fact that the learning signal from the dataset is 
richer per datapoint. Specifically, each graph contains around 150 nodes and 
several thousands of edges. Essentially, each graph contributes to learning the edge and node functions in proportion to
the number of edges and nodes in the graph, not as a single datapoint\footnote{This observation motivates augmenting the dataset with more random graphs even for the same experiment, not a single one as it is the case in this work. This was not attempted due to resource constraints but this investigation may form part of future works.}.
\section{Results}
\begin{figure}[ht!]
  \center
  \includegraphics[width=\textwidth]{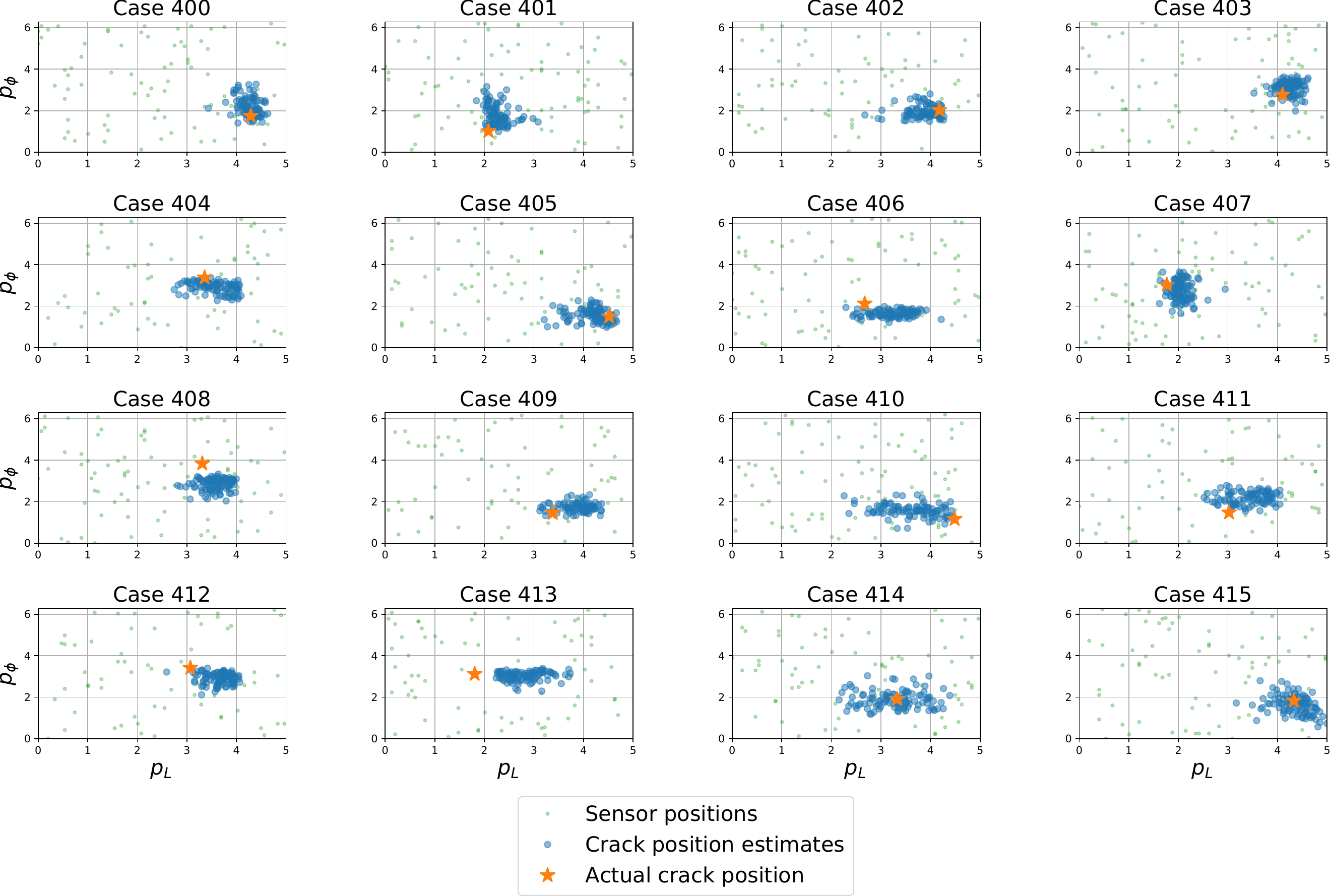}
  \caption{Strain-based localization results on test-set for some random cases for the \textbf{Bayesian GNN} model. The localization is performed on time-windows of 200 observations (250ms).
  Predictions are denoted in blue. The scatter on the predictions, is partially because of the Bayesian neural network layers and partially because of the different inputs for each segment.
  It is expected that for some time instants the localization is inaccurate since the crack does not contribute to the strain field surrounding it and cannot be detected.}
  \label{fig:localizationres}
\end{figure}

\begin{figure}[ht!]
  \center
  \includegraphics[width=\textwidth]{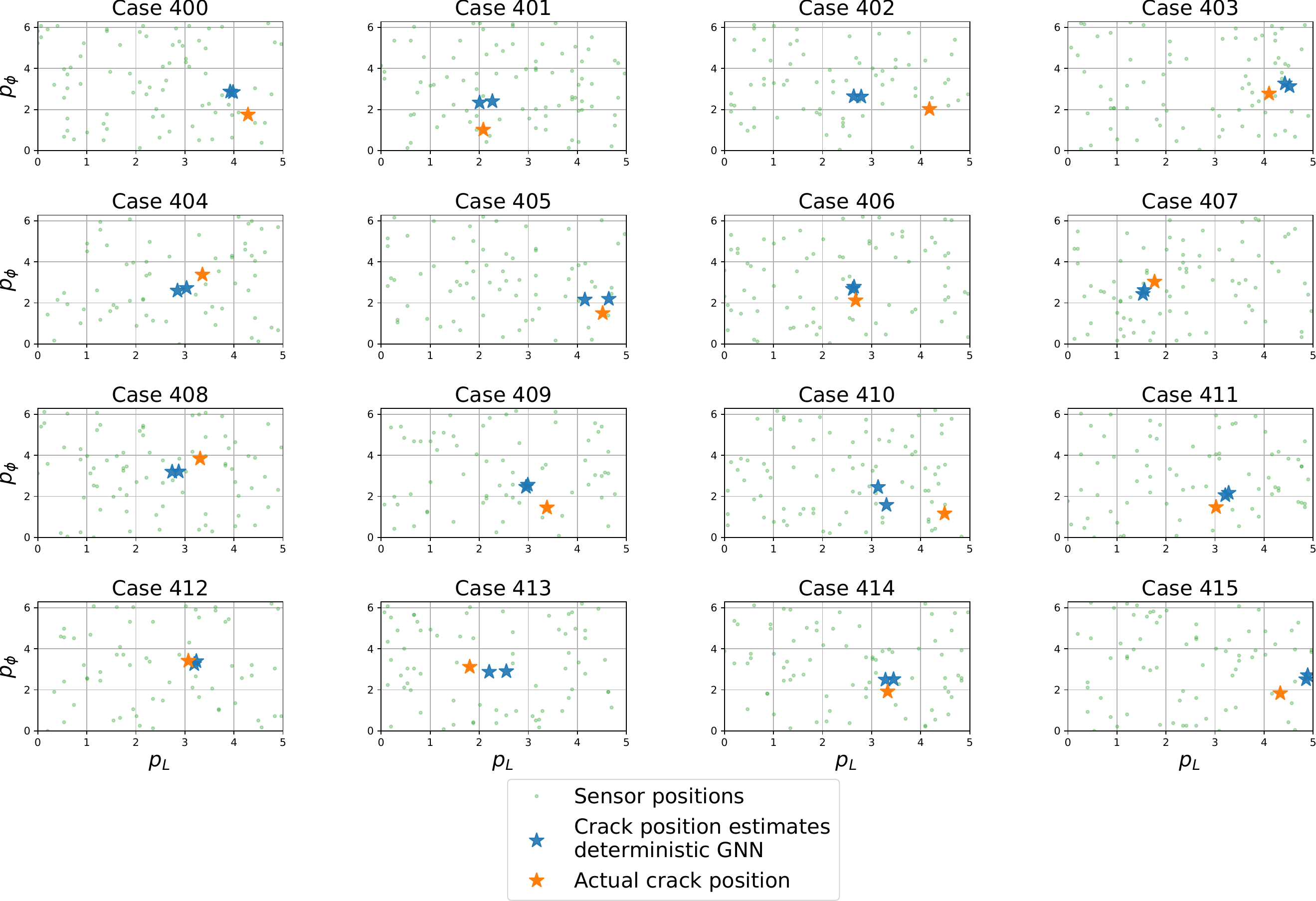}
  \caption{Strain-based localization results on test-set for some random cases for the \textbf{least-squares GNN} model. The localization is performed
  on two non-overlapping time-windows of 200 observations (250ms).
  Predictions are denoted in blue. The network returns deterministic and incorrect predictions with confidence.}
  \label{fig:detgnn}
\end{figure}

\begin{figure}[ht!]
  \center
  \includegraphics[width=\textwidth]{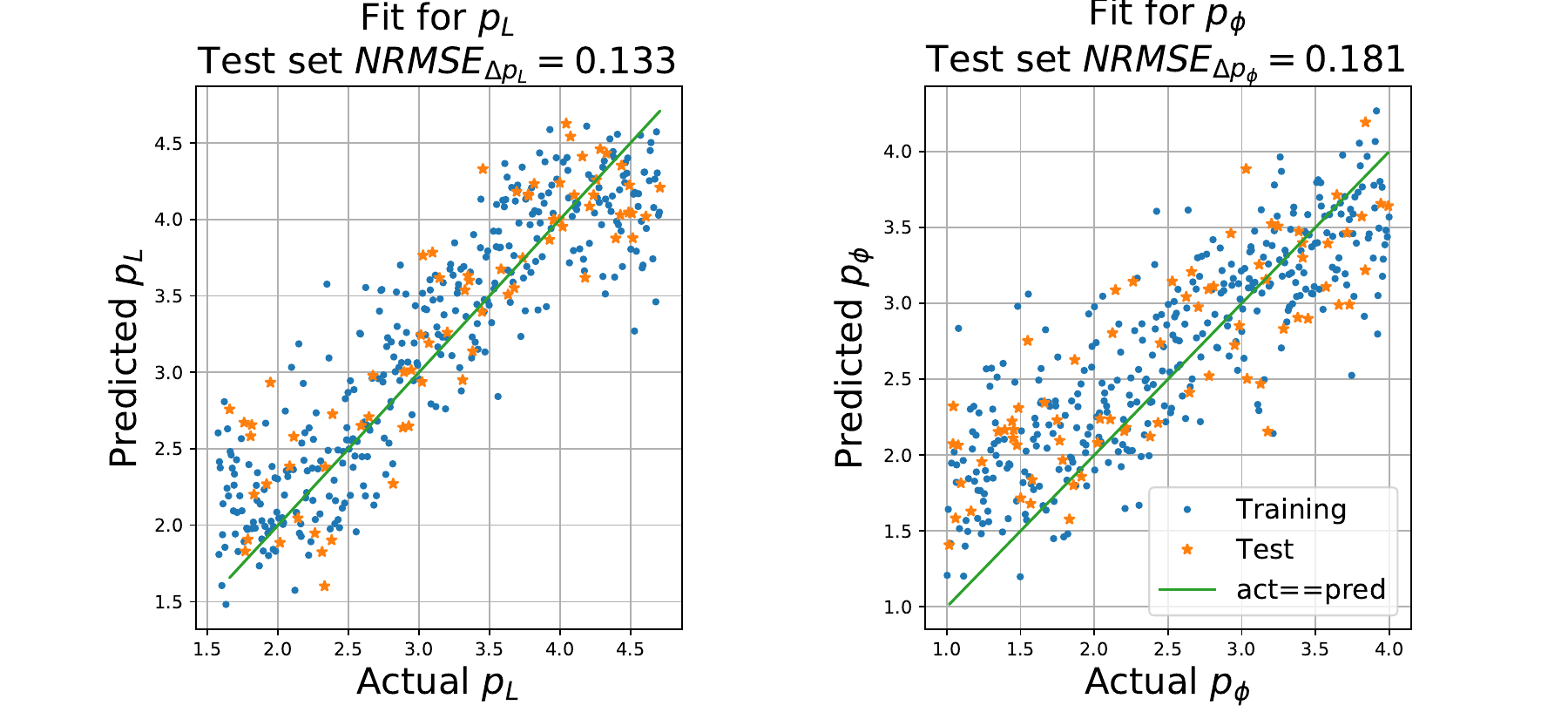}
  \caption{Strain-based localization results - all training and validation sets.}
  \label{fig:errs}
\end{figure}

The strain localization results for some random cases from the test-set are given in \autoref{fig:localizationres}. The GNN predicts 
a set of points close to the actual node position. The results on the angle of the crack were less accurate than the position estimates.
Because of the incorporation of Bayesian neural network layers, every forward estimation of the GNN the outputs are different, and each evaluation 
can be thought of as corresponding to a different model sampled from the posterior distribution of GNN models that was learned from the data. 
The Normalized Root Mean-Squared Error $L^{NRMSE}$, computed as,
\begin{equation}
  L^{NRMSE}(y_{pred} , y_{actual}) = \frac{\sqrt{\frac{\sum^N(y_{pred} - y_{actual})^2}{N}}} {\max(y_{act})-\min(y_{act})}
\end{equation}
is given together with all the training and test set examples in \autoref{fig:errs}. For comparison, localization results on the test-set for a deterministic GNN trained on least-squares loss, with the BNN layers replaced with simple FFNN layers, is given in \autoref{fig:detgnn}. Although the results are relatively close to the correct predictions, no estimates on the uncertainty of the predictions can be obtained.
\vspace{1cm}
\section{Conclusions}
In this work, graph neural networks with temporal convolution feature extractors were employed for a defect localization problem, using arbitrary configurations of sensors.
Because of the end-to-end differentiability of GNNs, future extensions of this work include optimal sensor placement optimization problems \cite{argyris2018bayesian}. 
The incorporation of data from multiple sensor types, such as active piezo-electric sensors can also be naturally implemented in the GNN framework.
In addition, GNN-based computation may be a viable alternative to Bayesian PDE inversion problems \cite{yang2020b}, where observations are sparse, in place 
of the grid-based CNN computational substrate. Finally, GNNs offer great potential for inference of data-driven Reduced Order Models (ROMs), directly from physical observations. GNNs have
already been successfully applied to learning complex physics with particle methods \cite{sanchez2020learning} and multi-body dynamics
\cite{sanchez2018graph} on simulation data.
\paragraph{Ackgnowledgement}
C. M. and E. C. would like to gratefully acknowledge the support of the European Research Council via the ERC Starting Grant WINDMIL (ERC-2015-
StG \#679843) on the topic of Smart Monitoring, Inspection and Life-Cycle Assessment of Wind Turbines. G. T. and K. W. were supported by the
EC Marie Skłodowska-Curie ETN grant “DyVirt” (764547).

\bibliographystyle{unsrt}
\bibliography{bibliography}

\begin{thebibliography}{10}

\bibitem{gulgec2019convolutional}
Nur~Sila Gulgec, Martin Tak{\'a}{\v{c}}, and Shamim~N Pakzad.
\newblock Convolutional neural network approach for robust structural damage
  detection and localization.
\newblock {\em Journal of Computing in Civil Engineering}, 33(3):04019005,
  2019.

\bibitem{mpnn}
Justin Gilmer, Samuel~S Schoenholz, Patrick~F Riley, Oriol Vinyals, and
  George~E Dahl.
\newblock Neural message passing for quantum chemistry.
\newblock {\em arXiv preprint arXiv:1704.01212}, 2017.

\bibitem{battaglia2018relational}
Peter~W Battaglia, Jessica~B Hamrick, Victor Bapst, Alvaro Sanchez-Gonzalez,
  Vinicius Zambaldi, Mateusz Malinowski, Andrea Tacchetti, David Raposo, Adam
  Santoro, Ryan Faulkner, et~al.
\newblock Relational inductive biases, deep learning, and graph networks.
\newblock {\em arXiv preprint arXiv:1806.01261}, 2018.

\bibitem{agathos2018multiple}
Konstantinos Agathos, Eleni Chatzi, and St{\'e}phane~PA Bordas.
\newblock Multiple crack detection in 3d using a stable xfem and global
  optimization.
\newblock {\em Computational mechanics}, 62(4):835--852, 2018.

\bibitem{agathos2020parametrized}
Konstantinos Agathos, St{\'e}phane~PA Bordas, and Eleni Chatzi.
\newblock Parametrized reduced order modeling for cracked solids.
\newblock {\em International Journal for Numerical Methods in Engineering},
  121(20):4537--4565, 2020.

\bibitem{mylonas2021machine}
Charilaos Mylonas.
\newblock {\em Machine Learning for Structural Health Assessment under
  Uncertainty: with applications in Wind Energy}.
\newblock PhD thesis, ETH Zurich, 2021.

\bibitem{gal2016dropout}
Yarin Gal and Zoubin Ghahramani.
\newblock Dropout as a bayesian approximation: Representing model uncertainty
  in deep learning.
\newblock In {\em International Conference on Machine Learning}, pages
  1050--1059, 2016.

\bibitem{kingma2015variational}
Durk~P Kingma, Tim Salimans, and Max Welling.
\newblock Variational dropout and the local reparameterization trick.
\newblock In {\em Advances in neural information processing systems}, pages
  2575--2583, 2015.

\bibitem{bayesbybackprop}
Charles Blundell, Julien Cornebise, Koray Kavukcuoglu, and Daan Wierstra.
\newblock Weight uncertainty in neural networks.
\newblock {\em arXiv preprint arXiv:1505.05424}, 2015.

\bibitem{lamb2020bayesian}
George Lamb and Brooks Paige.
\newblock Bayesian graph neural networks for molecular property prediction.
\newblock {\em arXiv preprint arXiv:2012.02089}, 2020.

\bibitem{dillon2017tensorflow}
Joshua~V Dillon, Ian Langmore, Dustin Tran, Eugene Brevdo, Srinivas Vasudevan,
  Dave Moore, Brian Patton, Alex Alemi, Matt Hoffman, and Rif~A Saurous.
\newblock Tensorflow distributions.
\newblock {\em arXiv preprint arXiv:1711.10604}, 2017.

\bibitem{vaepaper}
Diederik~P Kingma and Max Welling.
\newblock Auto-encoding variational bayes.
\newblock {\em arXiv preprint arXiv:1312.6114}, 2013.

\bibitem{argyris2018bayesian}
Costas Argyris, Sharmistha Chowdhury, Volkmar Zabel, and Costas Papadimitriou.
\newblock Bayesian optimal sensor placement for crack identification in
  structures using strain measurements.
\newblock {\em Structural Control and Health Monitoring}, 25(5):e2137, 2018.

\bibitem{yang2020b}
Liu Yang, Xuhui Meng, and George~Em Karniadakis.
\newblock B-pinns: Bayesian physics-informed neural networks for forward and
  inverse pde problems with noisy data.
\newblock {\em arXiv preprint arXiv:2003.06097}, 2020.

\bibitem{sanchez2020learning}
Alvaro Sanchez-Gonzalez, Jonathan Godwin, Tobias Pfaff, Rex Ying, Jure
  Leskovec, and Peter~W Battaglia.
\newblock Learning to simulate complex physics with graph networks.
\newblock {\em arXiv preprint arXiv:2002.09405}, 2020.

\bibitem{sanchez2018graph}
Alvaro Sanchez-Gonzalez, Nicolas Heess, Jost~Tobias Springenberg, Josh Merel,
  Martin Riedmiller, Raia Hadsell, and Peter Battaglia.
\newblock Graph networks as learnable physics engines for inference and
  control.
\newblock {\em arXiv preprint arXiv:1806.01242}, 2018.

\end{thebibliography}

\end{document}